# Comment on "New method for studying neutrino mixing and mass differences"

Murray Peshkin*

*Physics Division, Argonne National Laboratory, Argonne, IL 60439*

## Abstract

It has been suggested that the temporal oscillations in the rate of decay of Pr ions reported in the GSI K-capture experiment arise from interference between different values of the momentum of the Pr ions. It is shown here that any such interference cannot contribute to the decay rate of those ions.

### 1. Absence of interference between different momentum states

It has been proposed [1] that experiments studying the behavior of a radioactive ion which decays by capture of a K-shell electron and emission of a neutrino [2] may provide information about neutrino mixing and mass differences. In that proposal, the two neutrino masses enable two momentum components in the wave function of the parent ion to feed the same final state. The interference between those two momentum components, which have different energies, leads to an oscillatory time dependence of the decay rate. The magnitude and frequency of the oscillation can then be used to obtain information about the neutrino masses and mixing angle.

In fact, no such interference between different momenta can be observed in an experiment that measures the total decay rate independently of where the decay takes place.

*Proof:*
The counting rate $R(t)$ in a detector of the decay is given by

$$R(t) = \frac{2\pi}{\hbar} Tr\{\rho(t) H_{int} Q H_{int}\} = \frac{2\pi}{\hbar} \int \langle \mathbf{p}|\rho(t)|\mathbf{p'}\rangle\langle \mathbf{p'}|H_{int} Q H_{int}|\mathbf{p}\rangle d^3\mathbf{p}\, d^3\mathbf{p'}, \qquad (1)$$

where $\rho(t)$ is the density matrix of the parent ions, $H_{int}$ is the decay interaction term in the Hamiltonian, and $Q$ is the projection on all those decayed states that the detector senses. The decay interaction $H_{int}$ is invariant under spatial translation. If $Q$ is also invariant under translation, then $\langle \mathbf{x'}|H_{int} Q H_{int}|\mathbf{x}\rangle$ is a function of $(\mathbf{x}-\mathbf{x'})$ but not of $(\mathbf{x}+\mathbf{x'})$ so $\langle \mathbf{p'}|H_{int} Q H_{int}|\mathbf{p}\rangle$ is equal to $\delta(\mathbf{p}-\mathbf{p'})$ times some function $F(\mathbf{p})$. Then

$$R(t) = \frac{2\pi}{\hbar} \int \langle \mathbf{p}|\rho(t)|\mathbf{p}\rangle F(\mathbf{p}) d^3\mathbf{p} \qquad (2)$$

and no interference between different $\mathbf{p}$ contributes to the decay rate.

For this proof to apply, it is not necessary for $Q$ to be invariant under all translations. It is enough that, as in the GSI experiment, the detector senses all decays anywhere within a spatial region that includes all places that are accessible to the parent ions.

This absence of interference effects between different **p** is independent of whether the parent ions are confined to a circular beam as in the GSI experiment, to a linear beam, or to a box. It is also independent of any external fields, stationary or time dependent, that may be present, and of coupling to other dynamical systems as in beam cooling. The decay rate in Eq.(2) depends upon $\rho$ at the time of the measurement only, not upon how it evolved.

Eq.(2) implies only that the rate of decay at time $t$ is not affected by any interference between different momenta. Interference between components of the wave function having the same momentum but different energies is always present in a radioactive parent because an unstable state is a coherent mixture of a range of masses $m$. In the usual case of exponential decay the density matrix elements $\langle \mathbf{p},m|\rho(t)|\mathbf{p},m'\rangle$ are proportional to two Breit-Wigner amplitudes times $\exp\{i(E-E')t\}$, where $E = \sqrt{m^2 + p^2}$ and $E' = \sqrt{m'^2 + p^2}$. It is the interference between different $E$ for the same **p** that gives the exponential decay. To modulate that exponential decay with an oscillation requires a model in which the product $H_{int}QH_{int}$ connects states of the same momentum but with the energy difference $E$-$E'$ equal to $\hbar$ times the oscillation frequency.

## 2. Acknowledgements

This work was supported by the U.S. Department of Energy, Office of Nuclear Physics, under Contract No. DE-AC02-06CH11357. I thank John P. Schiffer for valuable suggestions.

*email: peshkin@anl.gov